\documentclass[12pt]{iopart}
\usepackage{upgreek,bbm,graphicx}

\def\hvarrho{\hat{\varrho}}
\def\hsigma{\hat{\sigma}}
\def\hD{\hat{D}}
\def\ha{\hat{a}}

\begin{document}

\title{Characterization of phase-averaged coherent states}

\author{ A.~Allevi$^{1,2}$, M.~Bondani$^{3,2}$, P.~Marian$^{4,5}$,
T.~A.~Marian$^{4}$ and S.~Olivares$^{6,7}$}

\address{$^{1}$Dipartimento di Scienza e Alta Tecnologia, Universit\`a degli Studi dell'Insubria,
      I-22100 Como, Italy}
\address{$^{2}$CNISM, UdR Como, I-22100 Como, Italy}
\address{$^{3}$Istituto di Fotonica e Nanotecnologie, C.N.R., I-22100 Como, Italy}
\address{$^{4}$Centre for Advanced  Quantum Physics, University of
     Bucharest, R-077125 Bucharest-M\u{a}gurele, Romania}
\address{$^{5}$Department of Physical Chemistry, University of Bucharest,
Boulevard Regina Elisabeta 4-12, R-030018  Bucharest, Romania}
\address{$^{6}$Dipartimento di Fisica, Universit\`a degli Studi di
     Milano, I-20133 Milano, Italy}
\address{$^{7}$CNISM, UdR Milano Statale, I-20133 Milano, Italy}

\eads{
\mailto{alessia.allevi@uninsubria.it},
\mailto{maria.bondani@uninsubria.it},\\
\mailto{paulina.marian@g.unibuc.ro},
\mailto{tudor.marian@g.unibuc.ro},\\
\mailto{stefano.olivares@fisica.unimi.it}
}

\begin{abstract}
  We present the full characterization of phase-randomized or
  phase-averaged coherent states, a class of states exploited in
  communication channels and in decoy state-based quantum key
  distribution protocols.  In particular, we report on the suitable
  formalism to analytically describe the main features of this class
  of states and on their experimental investigation, that results in
  agreement with theory. We also show the results we obtained by
  manipulating the phase-averaged coherent states with linear optical
  elements and testify their good quality by employing some
  non-Gaussianity measures and the concept of mutual information.
\end{abstract}

\pacs{42.50.Dv, 42.50.Ar, 03.65.Wj, 03.67.-a, 85.60.Gz}

% \submitto{\NJP}

\maketitle

\section{Introduction}\label{s:intro}
Laser radiation, which can be described in terms of coherent states,
plays a relevant role in practical communication schemes. One of the
main advantages of coherent states over more exotic quantum states,
such as squeezed ones, is that they can propagate over long distances,
only suffering attenuation and without altering their fundamental
properties. A coherent state is characterized by a Poissonian
photon-number statistics and a well-defined optical phase. Thus, one
can easily implement phase-shifted keyed communication in which the
logical information (the bit) is encoded in two coherent states with
the same amplitude and a $\pi$-difference in phase. Nevertheless, this
kind of communication channel lacks security. Remarkably, very
recently quantum key distribution involving coherent states and decoy
states has been realized and it has been pointed out that
phase-averaged coherent states may enhance the security of the channel
\cite{PRL:05,APL:07,EPJD}. In this case, the high degree of accuracy
in the phase randomization process is one of the main requirements.
\par
By contrast to coherent states, which are described by Gaussian
Wigner functions, phase-averaged coherent states clearly exhibit
non-Gaussian features \cite{oli:rev}.  Thus, the systematic study of
the nature of these states and the possibility to manipulate them can
be considered of real interest in enhancing the performances of the
communication protocols in which they are employed \cite{curty:09}.
\par
In this paper we investigate the main features of phase-averaged coherent
states and report on their fully experimental characterization by
addressing the measurement of the photon-number statistics and the
reconstruction of the Wigner function. Furthermore, we perform basic
manipulation experiments by means of linear optical elements, in order
to assess the usefulness of these states for communication and
information processing.  The detection of such states is performed in
the mesoscopic photon-number regime by means of hybrid photodetectors.
\par
The plan of the paper is as follows. In section~\ref{s:prcs} we recall
several properties of phase-averaged coherent states like
photon-number statistics and purity. We also present the experimental
scheme used for the generation, characterization and manipulation of
such states. Section~\ref{s:basics} is devoted to their basic
description using the characteristic functions and the corresponding
quasiprobability densities. In this section we also report on the
strategy and realization of the experimental reconstruction of the
Wigner function. Non-Gaussianity of the phase-averaged coherent states
and its experimental measurement are addressed in
section~\ref{s:nonG}. Section~\ref{s:advanced} investigates the linear
operations with phase-averaged coherent states performed by a beam
splitter. Here we give a complete analytic description of the output
one-mode reduced states. These superpositions turn out to be
Fock-diagonal and are interesting for quantum information processing.
We find a good agreement between theory and our experimental results
for what concerns non-Gaussianity and mutual information of the
beam-splitter output states. Our concluding remarks are drawn in
section~\ref{s:concl}.

\section{Quantum description of phase-averaged coherent
  states}\label{s:prcs}
A single-mode phase-randomized or phase-averaged coherent state (PHAV)
is obtained by randomizing the phase $\phi$ of a coherent state:
\begin{equation}
|\beta\rangle = \exp{\left(-\frac{1}{2}|\beta|^2\right)} \sum_{n=0}^{\infty}
\frac{|\beta|^n\, e^{i n \phi}}{\sqrt{n!}}
| n \rangle,
\end{equation}
with $\beta = |\beta|\,e^{i \phi}$. Any PHAV $\hvarrho$ is diagonal in
the photon-number basis, namely:
\begin{equation}\label{dm}
\hvarrho = \int_{0}^{2\pi} \frac{d\phi}{2\pi}\, | \beta \rangle\langle
\beta | = \sum_{n=0}^{\infty} \varrho_{nn}| n \rangle\langle n |,
\end{equation}
where
\begin{equation}\label{rho_nn}
\varrho_{nn}=\exp{\left(-|\beta|^2\right)}\frac{|\beta|^{2n}}{n!},
\end{equation}
is a Poissonian distribution.  Therefore, randomizing the phase of a
coherent state does not change its photon-number distribution [see
(\ref{rho_nn})]. Moreover, due to the diagonal structure of its
density matrix, the statistical properties of a PHAV can be fully
described by the photon-number distribution. Indeed, while
$|\beta\rangle$ is a pure state, the degree of purity $\mu[\hat
\varrho]$ of the PHAV $\hat \varrho$ can be directly evaluated by the
photon-number distribution and is given by:
\begin{eqnarray}\label{dp}
\mu[\hat \varrho]=\sum_{n=0}^{\infty}\varrho_{nn}^2=
\exp{\left(-2|\beta|^2\right)}\, I_0(2|\beta|^{2}),
\end{eqnarray}
where $I_0(z)$ is the modified zeroth-order Bessel function of the
first kind. The purity $ \mu[\hat \varrho]$ is a strictly decreasing
function of the mean number of photons $\langle\hat a^{\dag} \hat
a\rangle = {\rm Tr}[\hat \varrho\, \hat a^{\dag}\hat a ]=|\beta|^{2}$
($\hat a$ is the annihilation operator and $[\hat a,\hat
a^{\dag}]=\hat {\mathbbm I}$).
\par
\begin{figure}[htbp]
\centering\includegraphics[width=8cm]{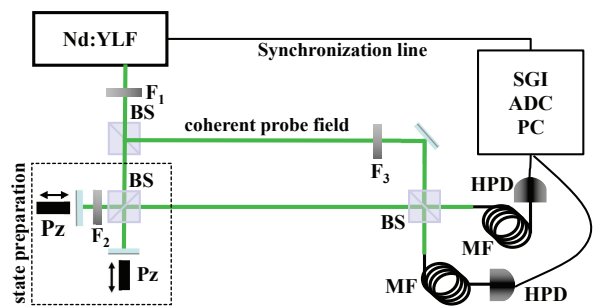}
\caption{Experimental setup. F$_j$: variable neutral density filter;
  BS: 50:50 beam splitter; Pz: piezoelectric movement; MF: multimode
  fiber (600 $\mu$m core); HPD: hybrid photodetector.} \label{setup}
\end{figure}

 
From the experimental point of view, we obtained this class of states
by sending the second-harmonics pulses of a mode-locked Nd:YLF laser
amplified at 500 Hz (High-Q Laser Production) to a mirror mounted on a
piezo-electric movement (see figure~\ref{setup}). The displacement of
the piezoelectric movement, which is controlled by a function
generator, is operated at a frequency of 100 Hz and covers a 12 $\mu$m
span \cite{ASL09}.
\par
\begin{figure}[htbp]
\centering\includegraphics[width=8cm]{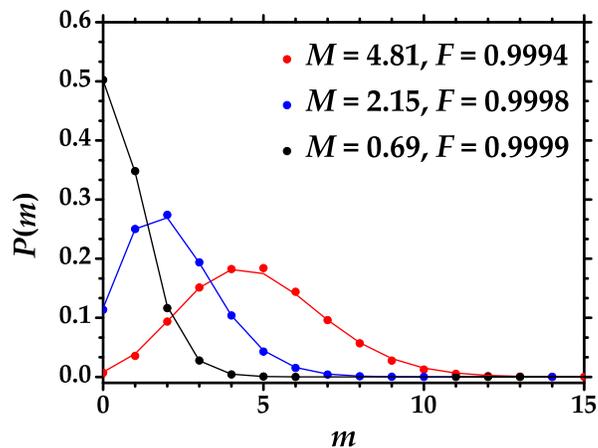}
\caption{Detected-photon distribution of PHAV state for three
  different mean values. Coloured dots: experimental data, lines:
  theoretical expectations.  The purity is $\mu[\hvarrho] = 0.13$ (red
  plot), $\mu[\hvarrho] = 0.20$ (blue plot) and $\mu[\hvarrho] = 0.39$
  (black plot).} \label{pnPHAV}
\end{figure}

In figure~\ref{pnPHAV} we show the detected photon distributions of
three PHAVs at different energy values, obtained by using a direct
detection scheme employing a hybrid photodetector (HPD, R10467U-40,
maximum quantum efficiency $\sim$ 0.5 at 500 nm, Hamamatsu)
characterized by a partial photon-counting capability and a linear
response up to 100 photons \cite{JMOself,andreoni09}.  In the same
figure, we also show the corresponding theoretical photon-number
statistics for detected photons [the photocount distribution is simply
obtained by using (\ref{rho_nn}) and replacing $|\beta|^2$ by $M
\equiv \eta|\beta|^2 $, where $\eta $ is the overall quantum
efficiency].
The good agreement between experimental data and theory can be
quantified by calculating the fidelity (see $F$ values reported in
figure~\ref{pnPHAV}): $F = \sum_{m=0}^{\bar{m}} \sqrt{P_{\rm th}(m)
  P(m)}$, in which $P_{\rm th}(m)$ and $P(m)$ are the theoretical and
experimental distributions, respectively, and the sum is extended up
to the maximum detected-photon number $\bar{m}$ above which both
$P_{\rm th}(m)$ and $P(m)$ become negligible.

\section{Basic characterization}\label{s:basics}
We insert (\ref{dm}) into the well-known definition of the
characteristic functions (CFs):
\begin{equation}
\chi({\lambda;s}) \equiv
\exp{\left(\frac{s}{2}|\lambda|^{2}\right)}\,
 {\rm Tr} [\hvarrho \hD(\lambda)],\quad (-1\leq s \leq 1)
\end{equation}
to write:
\begin{equation}\label{series:6}
\chi(\lambda;s)=\exp{\left(-\frac{1-s}{2}|\lambda|^{2}\right)}\,
\sum_{n=0}^{\infty}\varrho_{nn}\,L_n\left(|\lambda|^{2}\right).
\end{equation}
The series (\ref{series:6}) is obtained by substitution of the
diagonal matrix elements of the displacement operator $\hat{D}(\lambda)
=\exp(\lambda \ha^\dag - \lambda^* \ha)$ in the photon-number basis,
namely:
\begin{eqnarray}\label{D:nn}
\langle n
|\hD(\lambda)|n\rangle = \exp\left(-\frac12 |\lambda|^2\right)
\,L_n\left(|\lambda|^{2}\right),
\end{eqnarray}
where $L_n(z)$ is a Laguerre polynomial.  By using the generating
function of the Laguerre polynomials \cite{1,2}:
\begin{equation}
\sum_{n=0}^{\infty}L_n(z)\frac{x^n}{n!}={\rm e}^x J_0 \left(2\sqrt{ x z}\right)
\label{a1}\end{equation}
where $J_0(z)$ is a Bessel function of the first kind, we get:
\begin{equation}
\chi(\lambda;s)=\exp{\left(-\frac{1-s}{2}|\lambda|^{2}\right)}\,
J_0 (2 |\beta||\lambda|).\label{cf}
\end{equation}
The Fourier transform of the CFs in (\ref{cf}) gives us the set of 
quasiprobability densities \cite{gla:69,cah:69}:
\begin{equation}
W(z;s)=\frac{1}{\pi}\int_{\mathbbm C}
{\rm d}^{2}{\lambda}\; \exp{(z\lambda^*-z^*
  \lambda)}\,\chi(\lambda;s), \label{ws}
\end{equation}
which explicitly read \cite{1,2}:
\begin{equation}
W(z;s)=
\frac{2}{1-s}\exp{\left[-\frac{2}{1-s}\left(|\beta|^2+|z|^2\right)\right]}\,
I_0\!\left(\frac{4|z||\beta|}{1-s}\right).
\label{ws2}
\end{equation}
\par
If we set $s=-1$ in (\ref{ws2}), we obtain the $Q$ function:
\begin{eqnarray}
Q(z)&=&\frac{1}{\pi} W(z;-1)\nonumber \\
&=& \frac{1}{\pi}\exp{\left[-\left(|\beta|^2+|z|^2\right)\right]} \,
I_0 \left(2|z||\beta|\right).
\end{eqnarray}
\par
The $P$ function retrieved for $s=1$ by employing asymptotic
expansions of Bessel functions is expressed in terms of Dirac's
$\delta$ distribution:
\begin{eqnarray}\label{p}
P(z)&=&W(z;1)\nonumber \\
&=&\frac{1}{\pi}
\delta \left( |z|^2-|\beta|^2\right)=\frac{1}{2\pi |\beta|}
\delta \left(|z|-|\beta|\right),
\end{eqnarray}
and  the last equality follows from the properties of the $\delta $
distribution. Equation (\ref{p}) shows us that the mixed state
obtained by averaging over the phase of a pure coherent state
preserves the important feature of being at the classicality threshold
(remember that the coherent states are the only pure states at the
classicality threshold).
\par
\begin{figure}[htbp]
\centering\includegraphics[width=8cm]{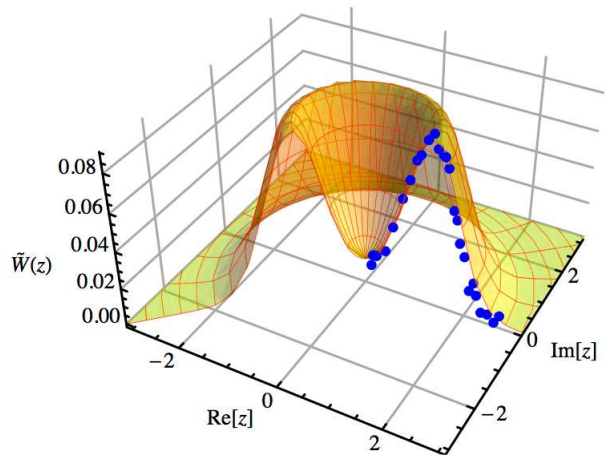}
\caption{Experimental reconstruction of a section of the Wigner
  function of a PHAV with $|\beta|^2 = 1.97$ and $\xi = 0.999$. Blue
  dots: experimental data, orange mesh: theoretical
  expectation.} \label{Wphav}
\end{figure}

Finally, for $s=0$ we get the Wigner function:
\begin{equation} \label{eq:wigPHAV}
W(z)=2
\exp{\left[-2\left(|\beta|^2+|z|^2\right)\right]}I_0\left(4|z||\beta|\right).
\end{equation}
which is positive everywhere in the phase space. Recently, states with
positive Wigner functions have become interesting for efficient
classical simulation of a broad class of quantum optics
experiments. In \cite{vei:12a,vei:12b} a
protocol for classical simulations using non-Gaussian states with
positive Wigner function was presented (see also the more recent
\cite{mari:PRL:12}). Note that the Wigner function in
(\ref{eq:wigPHAV}) is not Gaussian, a feature that becomes evident
from the plot of the function that shows a dip at the origin of the
phase space (see figure~\ref{Wphav}).
\par
In order to experimentally reconstruct the Wigner function of PHAVs we
adopted the same strategy presented in \cite{OLwigner} and based
on the measurements of the statistics of the state under investigation
mixed at a beam splitter (BS) with a coherent probe field whose
amplitude and phase can be continuously changed
\cite{wallentowitz96,banaszek96,all:PRA:09}. As the PHAV is a diagonal
state, its Wigner function is phase insensitive, $i.e.$ it exhibits a
rotational invariance about the origin of the phase space.
For this reason, in figure~\ref{Wphav} we show the experimental data
(blue dots) corresponding to a section of the Wigner function
superimposed to the theoretical surface (orange mesh):
\begin{equation}
\tilde{W}\left(\sqrt{\xi} \alpha\right) =
W\left(\sqrt{\xi}\alpha\right)
\exp\left[-(|\alpha|+|\beta|) \sqrt{1-\xi}\right], \label{eq:wignerPHAVoverlap}
\end{equation}
$\xi$ being the overall (spatial and temporal) overlap between the
probe field and PHAV \cite{OLwigner}.  In
(\ref{eq:wignerPHAVoverlap}), $|\beta|^2$ is now the mean number
of photons we measured, which includes the quantum efficiency.  In
fact, it is worth noting that for classical states the functional form
of the Wigner function is preserved also in the presence of losses and
its expression, given in terms of detected photons, reads $
\tilde{W}(\alpha) =\frac{2}{\pi} \sum_{m=0}^\infty (-1)^m
p_{m,\alpha}^{\ el}$, where $p_{m,\alpha}^{\ el}$ represent the
detected-photon-number distributions of the state to be measured displaced by
the probe field \cite{OLwigner}.

\section{Non-Gaussianity of PHAVs}\label{s:nonG}
In order to quantify the non-Gaussanity of PHAVs, here we compare
three different measures of non-Gaussianity recently introduced
\cite{ng:geno:07,ng:geno:08,ng:geno:10,ng:marian:12} and completely
determined by the density matrix in (\ref{dm}) and
(\ref{rho_nn}).  All the three measures compare the properties of the
state under investigation $\hvarrho$ with that of a Gaussian reference
state, $\hsigma$, having the same mean value and covariance matrix as
$\hvarrho$.  In the case of PHAVs, the reference Gaussian state is a
thermal state with mean occupancy $|\beta|^2$.
\par
The first measure is based on the Hilbert-Schmidt distance:
\begin{equation}\label{nonGA}
\varepsilon_{\rm A}[\hvarrho]
=\frac{D^2_{\rm HS}[\hvarrho, \hsigma]}{\mu[\hvarrho]}
=\frac{\mu[\hvarrho]+\mu[\hsigma]
-2\kappa[\hvarrho, \hsigma]}{2\mu[\hvarrho]},
\end{equation}
where $\mu$ is the purity of the state and $\kappa[\hvarrho,
\hsigma]=$Tr$[\hvarrho \hsigma]$ \cite{ng:geno:07}. The
Hilbert-Schmidt distance can be analytically calculated by using the
purity (\ref{dp}), the degree of purity of the reference thermal state
$\mu[\hat \sigma]=(2 |\beta|^2+1)^{-1}$ and the expression of
$\kappa[\hvarrho, \hsigma]$:
\begin{equation}
\kappa[\hvarrho, \hsigma]
=\frac{1}{|\beta|^{2}+1}\exp{\left(-\frac{|\beta|^{2}}{|\beta|^{2}+1}\right)}.
\end{equation}
\par
The second measure is the relative entropy of non-Gaussianity defined as:
\begin{equation}\label{nonGB}
\varepsilon_{\rm B}[\hvarrho] = S(\hsigma) - S(\hvarrho),
\end{equation}
where $S(\hvarrho) = -\hbox{Tr} [\hvarrho \ln \hvarrho]$ is the von
Neumann entropy of the state $\hvarrho$ \cite{ng:geno:08}. For all diagonal states, 
we have $S(\hvarrho)=-\sum_n \varrho_{nn}\ln\varrho_{nn}$,
where in the present case $\varrho_{nn}$ is given in (\ref{rho_nn}), and
$S(\hsigma)=\left( |\beta|^2 +1\right) \ln \left( |\beta|^2 +1\right) -
|\beta|^2 \ln|\beta|^2$.
\par
The last measure we study has been recently introduced in
\cite{ng:marian:12} and is based on the quantum fidelity, namely:
\begin{equation}
\varepsilon_{\rm C}[\hvarrho] = 1 -\sqrt{{\cal F}(\hvarrho,\hsigma)} ,
\end{equation}
where:
\begin{equation}
{\cal F}(\hvarrho,\hsigma)
= \left\{{\rm Tr}\left[\sqrt{\sqrt{\hvarrho}\,\hsigma\,\sqrt{\hvarrho}} \right] \right\}^2
\end{equation}
is the Uhlmann fidelity \cite{uhl:fid}. This measure can readily be
evaluated for Fock-diagonal states since they commute with their
reference thermal states.
\par
It is worth noting that the evaluation of the considered measures
only requires quantities that can be experimentally accessed by
direct detection, as they can be expressed in terms of
photon-number distributions \cite{OE12}.
\par
\begin{figure}[htbp]
\centering\includegraphics[width=8cm]{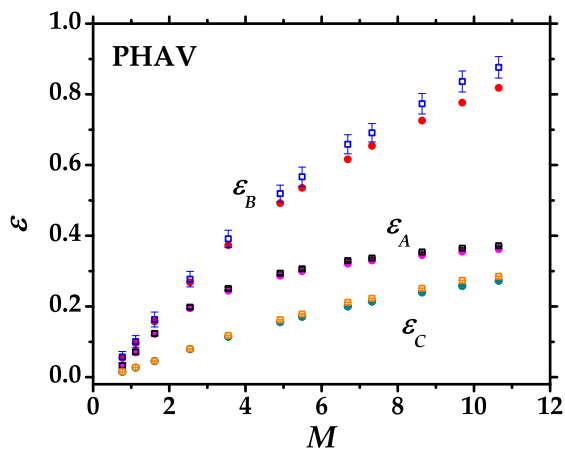}
\caption{Comparison between the three measures of non-Gaussianity in
  the case of PHAVs as functions of the mean number of photons. Open
  squared symbols: experimental data, dots: theoretical
  expectations.} \label{nonGphav3m}
\end{figure}
\begin{figure}[htbp]
\centering\includegraphics[width=8cm]{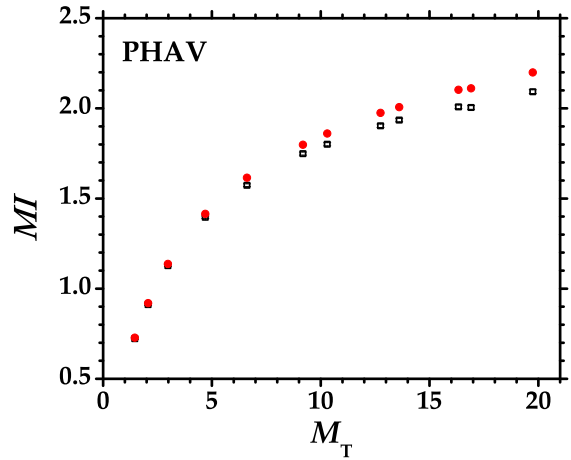}
\caption{Mutual information between the two outputs of the BS at which
  a PHAV with energy $M_{\rm T}$ is divided. Open squared symbols:
  experimental data, red dots: theory. The error bars are smaller than
  the symbol size.} \label{grMI2802}
\end{figure}
Figure~\ref{nonGphav3m} shows the behaviour of the three non-Gaussianity
measures as functions of the average number $M$ of detected photons in
the case of PHAVs.  It is evident that the behaviours of the three
measures are very similar to each other except for the absolute values
\cite{IJQInonG}.

\section{Advanced characterization and manipulation}\label{s:advanced}
When a coherent state $|\beta\rangle$ is mixed with the vacuum at a BS
with transmissivity $\tau$, the two emerging beams are excited in the
product state $|\sqrt{\tau}\beta\rangle \otimes
|\sqrt{1-\tau}\beta\rangle$ and thus are uncorrelated. Nevertheless,
when we consider a PHAV as the input state, a correlation arises at
the two outputs, even if intensity correlations still vanish
\cite{why1}. The total amount of correlation of the output bipartite state
$\hvarrho_{12}$ can be evaluated in terms of the mutual information
($MI$):
\begin{equation}\label{MI}
MI(\hvarrho_{12}) = S(\hvarrho_1)+S(\hvarrho_2) - S(\hvarrho_{12}),
\end{equation}
where $\hvarrho_k = {\rm Tr}_h[\hvarrho_{12}]$ ($h,k=1,2$ and $k \ne
h$) are the output reduced states and $S(\hvarrho)$ is the von Neumann
entropy. In figure~\ref{grMI2802} we plot the experimental data (open
squared symbols) and the theoretical predictions (red dots) of the
$MI$ as a function of the energy of the input PHAV.  We experimentally
measured this parameter by using a scheme involving two hybrid
photodetectors to simultaneously detect the light at the two outputs
of the BS \cite{OLcorr}, as shown in figure~\ref{setup}.  Since from the
experimental point of view it is not possible to measure both the
input and output states simultaneously, in order to assess the last
term in (\ref{MI}) we assumed that the input state was a PHAV with
energy equal to the sum of the two output channels (losses at the BS
are negligible). We also notice that in figure~\ref{grMI2802} the slight
discrepancy between experiment and theory appearing at increasing
values of the input energy can be due to some saturation effect of the
acquisition chain.
\par
We have also investigated another interesting state obtained by the
interference of two PHAVs (see figure~\ref{setup}): we will refer to
this state, which is still diagonal in the photon-number basis, as
2-PHAV \cite{OE12}. The 2-PHAV can find useful applications in passive
decoy state quantum key distribution \cite{curty:09}.  To describe the
2-PHAV state, we start observing that when two uncorrelated field
modes described by a product CF $\chi_0(\lambda_1, \lambda_2;s)=
\chi_{01} (\lambda_1;s)\, \chi_{02}(\lambda_2;s)$ are mixed at a BS
with transmissivity $\tau$, the output two-mode CF may be written as
follows $\chi(\lambda_1, \lambda_2;s)=\chi_{01} (\zeta_1;s) \,
\chi_{02}(\zeta_2;s)$ with $\zeta_1=\lambda_1
\sqrt{\tau}-\lambda_2\sqrt{1-\tau},\;\; \zeta_2=\lambda_2
\sqrt{\tau}+\lambda_1\sqrt{1-\tau}$ \cite{oli:rev}.  Therefore, the CF
of the 2-PHAV state, obtained by taking only one of the output modes,
can be formally written as (the CF of the other mode can be obtained
by replacing $\tau$ with $1-\tau$):
\begin{equation}
\chi(\lambda;s)=\chi_{01}\left(\lambda\sqrt{\tau};s\right)\,
\chi_{02}\left (\lambda\sqrt{1-\tau};s\right),\label{cf1}
\end{equation}
which follows from the partial-trace rule in the reciprocal phase
space. (\ref{cf1}) gives the following multiplication rule for the
input states of the type (\ref{cf}):
\begin{eqnarray}
\chi(\lambda;s)=\exp{\left(-\frac{1-s}{2}|\lambda|^{2}\right)}
% \nonumber\\ \times
J_0 \left(2 |\beta_1||\lambda|\sqrt{\tau}\right)
J_0 \left(2 |\beta_2||\lambda|\sqrt{1-\tau}\right),\label{cf2}
\end{eqnarray}
$\beta_1$ and $\beta_2$ are the coherent amplitudes of the
corresponding interfering PHAVs.  Note that $\chi(\lambda;s) \equiv
\chi(|\lambda|;s)$, as expected for phase-insensitive states.
\par
In order to obtain the photon statistics of the 2-PHAV, we can follow
two strategies. On the one hand, we can use the expansion of the
density operator in terms of displacement operators \cite{gla:69}:
\begin{equation}
\hvarrho = \frac{1}{\pi}\int_{\mathbbm C} {\rm d}^2\lambda\, \chi(\lambda;0)\,\hD(-\lambda).
\end{equation}
Therefore, the matrix elements of a phase-insensitive single-mode state
described by the CF $\chi(|\lambda|) \equiv \chi(\lambda;0)$ may be
written as:
\begin{equation}
\varrho_{lm}=2\;\delta_{lm}\int_0^{\infty} {\rm d}|\lambda|\,
|\lambda|\exp{\left(-\frac{1}{2}|\lambda|^{2}\right)}\,
\chi(|\lambda|)\, L_m\left(|\lambda|^{2}\right),\label{ro1}
\end{equation}
where we used (\ref{D:nn}) and performed the integration over the
polar angle, thus being left with an integral over $|\lambda|$.
\par
On the other hand, we can exploit high-order correlation functions
via the series \cite{gla:69,cah:69}:
\begin{equation}
\varrho_{nn}=\frac{1}{n!}\sum_{k=0}^{\infty}\frac{(-1)^{k-n}}{(k-n)!}
\langle (\ha^{\dag})^k \ha^k\rangle.\label{pnd3}
\end{equation}
Using a generating-function method, we are able to derive the
$k$th-order correlation functions in terms of Legendre polynomials
$P_k(x)$:
\begin{equation}
\langle (\ha^{\dag})^k \ha^k\rangle= \left[|\beta_1|^2 \tau+|\beta_2|^2 (1-\tau)\right]^k\,
u^k \, P_k\left(\frac{1}{u}\right),
\label{hcf}\end{equation}
where $u$ is the ratio:
\begin{equation} 
u=\frac{\big| |\beta_1|^2 \tau -|\beta_2|^2 (1-\tau)\big|
}{|\beta_1|^2 \tau+|\beta_2|^2 (1-\tau)}.
\label{udef}
\end{equation}
The $k$th-order normalized correlation functions are then:
 \begin{equation}
g^{(k)}(0)=\frac{\langle (\ha^{\dag})^k \ha^k\rangle}{\langle \ha^{\dag}
  \ha\rangle^k}=u^kP_k\left(\frac{1}{u}\right).
\label{nhcf}\end{equation}
Now, thanks to the relationship (\ref{pnd3}) we obtain the density
matrix elements as series expansions involving Legendre polynomials:
\begin{eqnarray}
\varrho_{nn}=\,\frac{1}{n!}\sum_{k=0}^{\infty}\frac{(-1)^{k-n}}{(k-n)!}
% \nonumber\\ \times
\left[|\beta_1|^2 \tau+|\beta_2|^2 (1-\tau)\right]^k\,
u^k\, P_k\left(\frac{1}{u}\right).
\label{pnd2}
\end{eqnarray}
In particular, if $\tau=1/2$ (balanced BS) and $|\beta_1|=|\beta_2| =
|\beta|$ (identical inputs), we find a simple result for the degree of
coherence (\ref{nhcf}), namely:
\begin{equation}\label{gk:bbs}
g^{(k)}(0)=\frac{(2k-1)!!}{k!} >1, \qquad (k>1),
\end{equation}
which indicates a super-Poissonian photon statistics. The
photon-number distribution is obtained after some algebra via
(\ref{pnd3}) as:
\begin{eqnarray}
\varrho_{nn}=\, \frac{(2n-1)!!}{(n!)^2}
% \nonumber\\ \times
\,|\beta|^{2n}\,\exp{\left(-2|\beta|^2\right)}
  \,{_{1}F_{1}}\left(\frac{1}{2}; n+1; 2|\beta|^2\right),\label{rho_nn_2p}
\end{eqnarray}
where ${_{1}F_{1}}(q; p; z)$ is a confluent hypergeometric function \cite{1,zambra07}.
\par
\begin{figure}[htbp]
\centering\includegraphics[width=8cm]{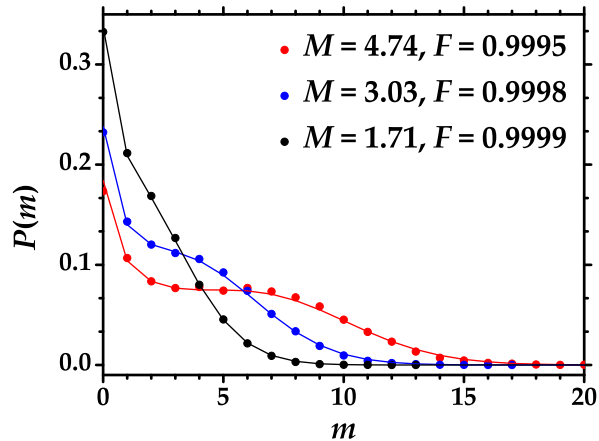}
\caption{"Detected-photon distribution of a balanced 2-PHAV state for
  three different mean values. Coloured dots: experimental data, lines:
  theoretical expectations (\ref{rho_nn_2p}).  The purity is $\mu[\hvarrho] = 0.09$ (red
  plot), $\mu[\hvarrho] = 0.13$ (blue plot) and $\mu[\hvarrho] = 0.21$
  (black plot).} \label{pn2PHAV}
\end{figure}
In figure~\ref{pn2PHAV} we have plotted the photon-number distribution
(32) for different energy values (coloured lines).  We remark the good
agreement between experimental data (coloured dots) and theory
predictions confirmed also by the high values of the fidelity.
\par
Finally, we have obtained the set of quasiprobability densities
associated with a 2-PHAV as having the following expansion:
\begin{eqnarray}
\fl W(\alpha;s) = \,2\exp{\left(-\frac{2 |\alpha|^2}{1-s}\right)}
\nonumber\\ \times
\sum_{k=0}^{\infty}\frac{(-1)^{k}}{k!}\left(\frac{2(|\beta_1|^2 T+|\beta_2|^2
    R)}{1-s}\right)^k
% \nonumber\\ \times
 u^k\, P_k\left(\frac{1}{u}\right)L_k
 \left(\frac{2|\alpha|^2}{1-s}\right),  \label{th:W2phav}
\end{eqnarray}
where $P_k(z)$ are Legendre polynomials and $u$ is defined in
(\ref{udef}).  In particular, the quasiprobability densities of
the balanced state (\ref{rho_nn_2p}) have a simpler form due to the explicit
correlation functions (\ref{gk:bbs}). We get:
\begin{eqnarray}
W(\alpha;s)=\,2\exp{\left(-\frac{2 |\alpha|^2}{1-s}\right)}
% \nonumber\\ \times
\sum_{k=0}^{\infty}\frac{(2k-1)!!}{(k!)^2}\left(-\frac{2|\beta|^2 }{1-s}\right)^k 
 L_k \left(\frac{2|\alpha|^2}{1-s}\right).
\end{eqnarray}
\par
\begin{figure}[htbp]
\centering\includegraphics[width=8cm]{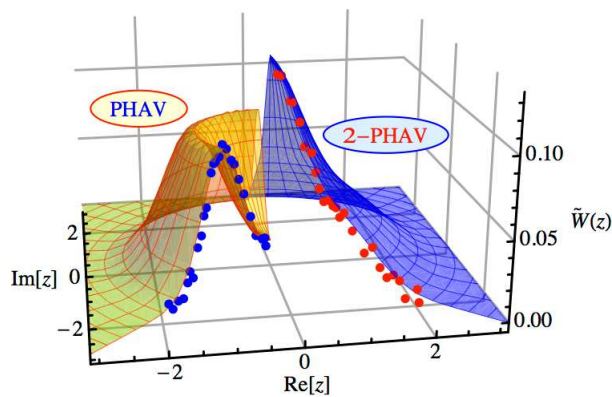}
\caption{Left: Experimental reconstruction of a section of the Wigner
  function of a PHAV with $|\beta|^2 = 1.97$ and $\xi = 0.999$.  Blue
  dots: experimental data, orange mesh: theoretical
  expectation. Right: Experimental reconstruction of a section of the
  Wigner function of a balanced 2-PHAV with $|\beta_1|^2 = 1.03$,
  $|\beta_2|^2 = 0.91$, $\xi_{\rm P} = 0.95$ and $\xi_{\rm S} = 1$.
  Red dots: experimental data, blue mesh: theoretical
  expectation.} \label{W2phav}
\end{figure}
\begin{figure}[htbp]
\centering\includegraphics[width=0.5\columnwidth]{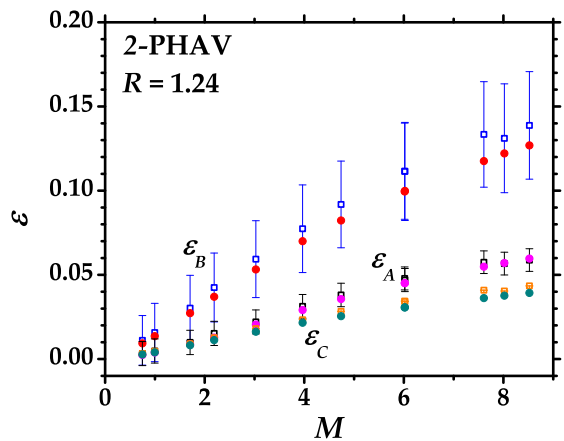}
\caption{Comparison between the three measures of non-Gaussianity in
  the case of balanced 2-PHAVs as functions of the mean number of
  photons. Open squared symbols: experimental data, dots: theoretical
  expectations.} \label{nonGvsmean3m}
\end{figure}
\begin{figure}[htbp]
\centering\includegraphics[width=8cm]{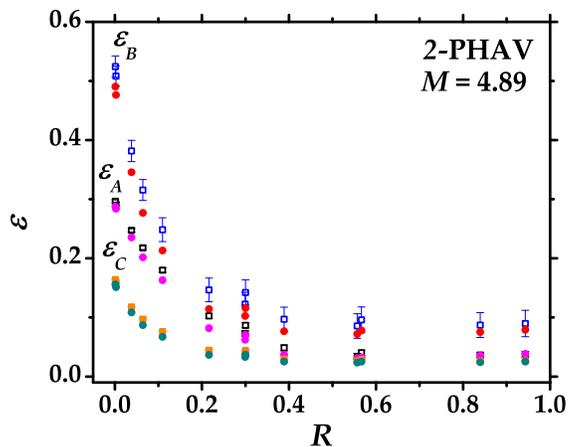}
\caption{Comparison between the three measures of non-Gaussianity in
  the case of 2-PHAVs at fixed total energy as functions of the
  balancing between the two single PHAVs.  Open squared symbols:
  experimental data, dots: theoretical
  expectations.} \label{gr2phav0805}
\end{figure}
\begin{figure}[htbp]
\centering\includegraphics[width=8cm]{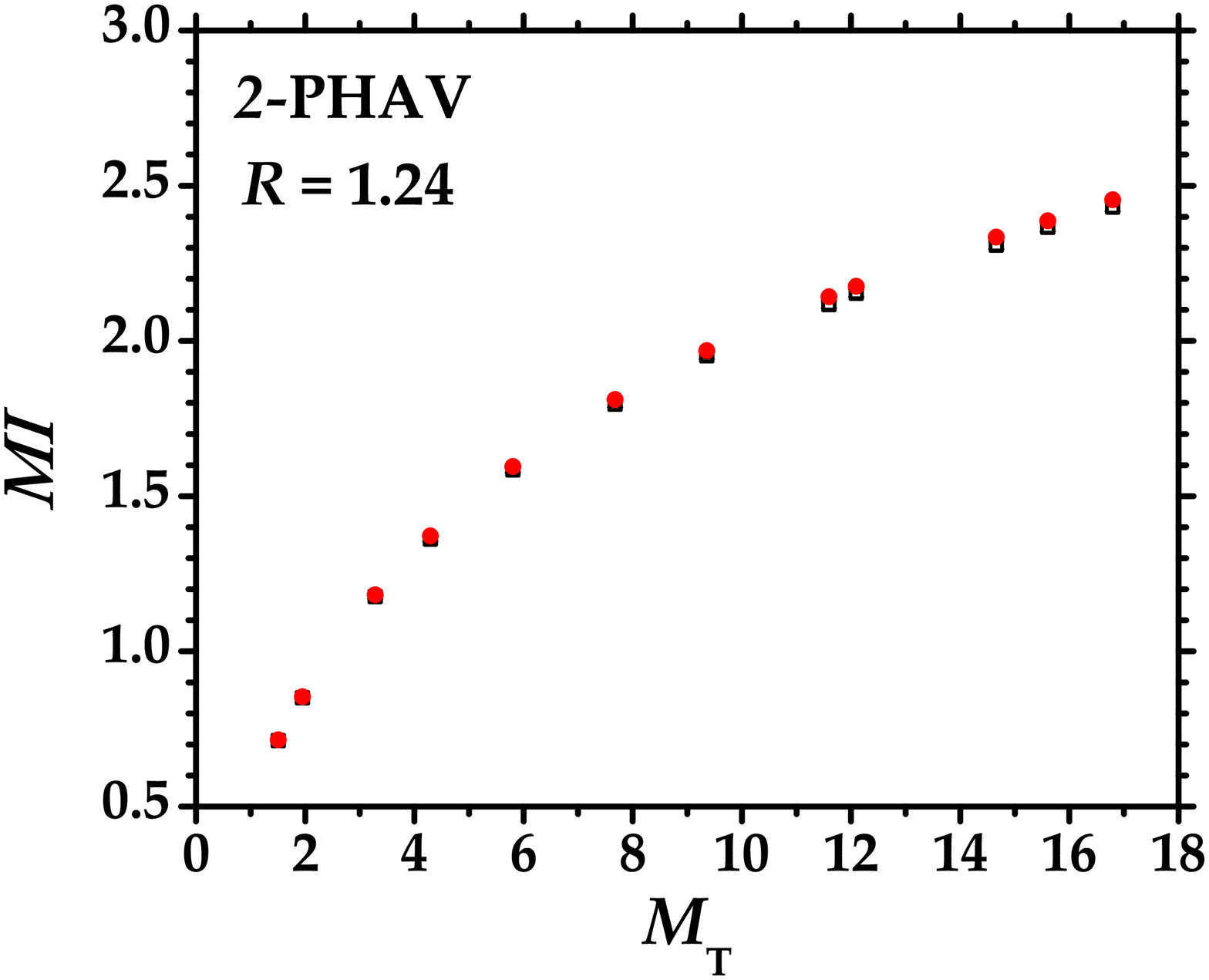}
\caption{Mutual information between the two outputs of the BS at which
  a balanced 2-PHAV is divided as a function of the total input energy
  $M_T$. Open squared symbols: experimental data, red dots:
  theory. The error bars are smaller than the symbol
  size.} \label{grMI0103}
\end{figure}
\begin{figure}[htbp]
\centering\includegraphics[width=8cm]{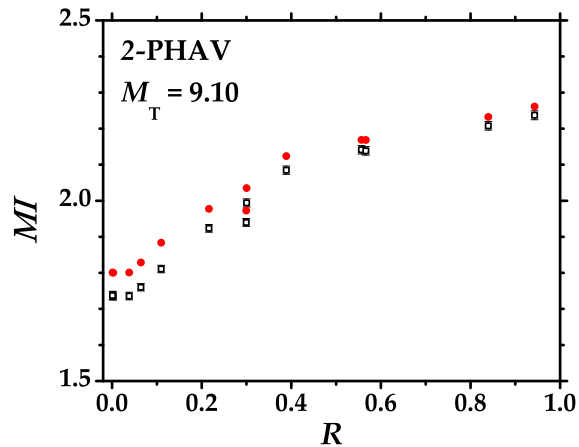}
\caption{Mutual information between the two outputs of the BS at which
  a 2-PHAV at fixed input energy $M_T$ is divided as a function of the
  ratio $R=|\beta_1|/|\beta_2|$ between the energies of the two PHAVs
  generating the 2-PHAV state. Open squared symbols: experimental
  data, red dots: theory.  The error bars are smaller than the symbol
  size.} \label{grMI0805}
\end{figure}

In figure~\ref{W2phav} we report a section of the phase-insensitive
Wigner function of a 2-PHAV (on the right) obtained by the
interference at a balanced BS of two identical PHAVs (whose section of
Wigner function is shown on the left). Even in the case of 2-PHAV, the
experimental data (red dots) are well superimposed to the theoretical
surface (blue mesh):
\begin{eqnarray}
\fl \tilde{W}_{\rm 2-PHAV}(\sqrt{\xi_{\rm P}} \alpha) =\,
W_{\rm 2-PHAV}(\sqrt{\xi}_{\rm P} \alpha)
% \nonumber \\ \times  
\exp\Big[-|\alpha| \sqrt{1-\xi_{\rm P}}
- (|\beta_1|+|\beta_2|) \sqrt{1-\xi_{\rm S}} \Big], \label{eq:wigner2PHAVoverlap}
\end{eqnarray}
where $\xi_{\rm P}$ describes the overall overlap between the probe
and the 2-PHAV and $\xi_{\rm S}$ the overall overlap between the two
components of the 2-PHAV.  It is evident that the single PHAV has a
dip at the origin of the phase space, whereas the 2-PHAV has a
peak. This difference results in a reduction of non-Gaussianity of the
2-PHAV with respect to that of a single PHAV at fixed energy, as
testified by the non-Gaussianity measures introduced above
\cite{OE12,IJQInonG}.  To stress this result, in
figure~\ref{nonGvsmean3m} we show the behaviour of the three measures
as functions of the energy values $M$ for a balanced 2-PHAV: we can
notice that the absolute values of $\varepsilon_k$, $k=A,B,C$ are
smaller than the ones we obtained in the case of a single PHAV.
Moreover, in figure~\ref{gr2phav0805} we plot the same measures as
functions of the balancing between the two components of the 2-PHAV at
fixed energy value $M$. As one may expect, the three measures
monotonically decrease at increasing the balancing. In fact, the most
unbalanced condition reduces to the case in which there is only a
single PHAV, whereas the most balanced one corresponds to have a
balanced 2-PHAV.
As in the case of a single PHAV, when a 2-PHAV state is mixed with the
vacuum at a 50:50 BS, the two outputs show a correlated nature,
testified by the non-zero value of the mutual information.  In
figure~\ref{grMI0103} we report the $MI$ between the two outputs of the
BS at which a balanced 2-PHAV is divided as a function of the input
energy value.  Furthermore, figure~\ref{grMI0805} shows the $MI$ at
fixed input energy of the 2-PHAV as a function of the ratio
$R=|\beta_1|/|\beta_2|$ between the two single PHAVs used to generate
the 2-PHAV state.
\par
In both figures the accordance between the experimental data (open
squared symbols), whose error bars are smaller than the symbol size,
and the theoretical predictions (red dots) is good. As in the case of
a single PHAV, in order to assess the last term in (\ref{MI}) we
again assumed that the input state was a 2-PHAV with energy equal to
the sum of the two output channels.

\section{Concluding remarks}\label{s:concl}
In conclusion, we have studied the main properties of PHAVs: we have
developed an analytic description and verified the theoretical
predictions by means of a direct detection scheme involving hybrid
photodetectors.  In detail, we have investigated the detected
photon-number distribution and the Wigner function that is
non-Gaussian. Moreover, we have used three different non-Gaussianity
measures, all based on quantities experimentally accessed by direct
detection, to quantify the non-Gaussianity amount and have proven the
consistency of the different approaches. Furthermore, we have
manipulated PHAVs by means of linear optical elements and generated a
new class of phase-randomized states, namely 2-PHAVs, obtained as
superpositions of two PHAVs at a BS. The consistent experimental and
theoretical results we have obtained in the characterization of both
PHAVs and their superpositions 2-PHAVs reinforce the possibility of
using them for applications to communication protocols. These
classical states appear to be robust, experimentally accessible and
theoretically convenient. The investigation of the non-Gaussianity and
correlations of some other output BS-states manipulated by conditional
measurements is one of our present interests.

\section*{Acknowledgments}
This work has been supported by MIUR (FIRB ``LiCHIS'' - RBFR10YQ3H)
and by the Romanian National Authority for Scientific Research through
Grant No.~PN-II-ID-PCE-2011-3-1012 for the University of Bucharest.


\section*{References}


\begin{thebibliography}{30}

\bibitem{PRL:05} Lo H-K, Ma X and Chen K 2005 \PRL {\bf 94} 230504

\bibitem{APL:07} Zhao Y, Qi B and Lo H-K 2007 {\it Appl. Phys. Lett.} {\bf 90} 044106

\bibitem{EPJD} Inamori H, L$\ddot{\rm u}$tkenhaus N and Mayers D 2007
  {\it Eur. Phys. J. D} {\bf 41} 599

\bibitem{oli:rev} Olivares S {\it Eur. Phys. J. Special Topics} 2012 {\bf 203} 3

\bibitem{curty:09} Curty M, Moroder T, Ma X and L\"{u}tkenhaus N 2009
  {\it Opt. Lett.} {\bf 34} 3238

\bibitem{ASL09} Bondani M, Allevi A and Andreoni A  2009 {\it Adv. Sci. Lett.} {\bf 2} 463 

\bibitem{JMOself} Bondani M, Allevi A, Agliati A and Andreoni A
  2009 {\it J. Mod. Opt.} {\bf 56} 226

\bibitem{andreoni09} Andreoni A and Bondani M 2009 {\it Phys. Rev. A} {\bf 80} 013819

\bibitem{1} Erd\'{e}lyi A, Magnus W, Oberhettinger F and  
Tricomi F G 1953 {\it  Higher Transcendental Functions}  (New
York, McGraw--Hill,  vol~1 and 2)
% :  Vol. 1 for hypergeometric functions and
% Legendre polynomials, Vol. 2 for Bessel functions and Laguerre
% polynomials.

\bibitem{2} Watson G N 1944 {\em A treatise on the theory of Bessel
    functions} (Cambridge University Press, Second edition)

\bibitem{gla:69} Cahill K E and Glauber R J 1969 {\it Phys. Rev.} {\bf 177} 1857

\bibitem{cah:69} Cahill K E and Glauber R J 1969 {\it Phys. Rev.} {\bf 177} 1882

\bibitem{vei:12a}  Veitch V, Ferrie C, Gross D and Emerson J 2012 {\it
    New J. Phys.} {\bf 14} 113011

\bibitem{vei:12b} Veitch V, Wiebe N, Ferrie C and Emerson J 2012
  Efficient simulation scheme for a class of quantum optics
  experiments with non-negative Wigner representation {\it Preprint}
  arXiv:1210.1783  [quant-ph]

\bibitem{mari:PRL:12} Mari A and Eisert J 2012 \PRL {\bf 109} 230503

\bibitem{OLwigner} Bondani M, Allevi A and Andreoni A  2009 {\it Opt. Lett.} {\bf 34} 1444

\bibitem{wallentowitz96} Wallentowitz S and Vogel W 1996 {\it Phys. Rev. A} {\bf 53} 4528

\bibitem{banaszek96} Banaszek K and W$\acute{\rm o}$dkiewicz K 1996
  \PRL {\bf 76} 4344

\bibitem{all:PRA:09} Allevi A, Andreoni A, Bondani A, Brida G,
  Genovese M, Gramegna M, Olivares S, Paris M G A, Traina P and
  Zambra G 2009 {\it Phys. Rev A} {\bf 80} 022114

\bibitem{ng:geno:07} Genoni M G, Paris M G A and Banaszek K 2007
  {\it Phys. Rev. A} {\bf 76} 042327

\bibitem{ng:geno:08} Genoni M G, Paris M G A and Banaszek K
  2007 {\it Phys. Rev. A} {\bf 78} 060303(R)

\bibitem{ng:geno:10}  Genoni M G and Paris M G {\it Phys. Rev. A} 2010
  {\bf 82} 052341.

\bibitem{ng:marian:12} Ghiu I, Marian P and Marian T A 2010 Measures
  of non-Gaussianity for one-mode field states  {\it Preprint}
  arXiv:1210.1929 [quant-ph]

\bibitem{uhl:fid} Uhlmann A 1976 {\it Rep. Math. Phys.} {\bf 9} 273

\bibitem{uhl:fid2}  Uhlmann A 1986 {\it Rep. Math. Phys.} {\bf 24} 229

\bibitem{OE12} Allevi A, Olivares S and Bondani M 2012 {\it Opt. Express}
  {\bf 20} 24850

\bibitem{IJQInonG} Allevi A, Olivares S and Bondani M 2012 {\it
    Int. J. Quantum Inf.} {\bf 10} 1241006

\bibitem{why1} This results follows from the form of the joint photon
  statistics of the two outgoing modes that is factorized and reads
  $p(n,m) = P(n;\tau |\beta|^2)\,P(m;(1-\tau)
  |\beta|^2)$, where $P(n;N) = \exp (-N)\,N^n(n!)^{-1}$ is the
  Poisson distribution.

\bibitem{OLcorr} Allevi A, Bondani M and Andreoni A 2012 {\it Opt. Lett.} {\bf 35} 1707  

\bibitem{zambra07} Zambra G, Allevi A, Bondani M, Andreoni A and Paris
  M G A 2007 {\it Int. J. Quantum Inf.} {\bf 5} 305


\end{thebibliography}
\end{document}